\documentclass{raa}
\usepackage{graphicx,times}             
\usepackage{natbib}
\usepackage{verbatim}
\PassOptionsToPackage{normalem}{ulem}
\usepackage{ulem}
\usepackage{amssymb,amsmath}
\usepackage{multirow}
\usepackage{booktabs}
\usepackage{color}
\usepackage{ulem}
\usepackage{threeparttable}
\bibpunct{(}{)}{;}{a}{}{,}
\usepackage{xcolor}

\usepackage[colorlinks=true,linkcolor=green,anchorcolor=red,citecolor=blue,filecolor=red,urlcolor=red]{hyperref}
\begin{document}
   \title{ASCNet: Research on all-sky camera images classification at the Muztagh-ata site
}

   \volnopage{Vol.0 (20xx) No.0, 000--000}      
   \setcounter{page}{1}          

   \author{Siqi Wang
      \inst{1,2}
   \and Qi Fan
      \inst{3}
   \and Wenbo Gu\thanks{*Corresponding author}
      \inst{1,2}
    \and Haozhi Wang
      \inst{1,2}
    \and AYZADA Jumahali
      \inst{1,2}
    \and Lixian Shen
      \inst{1,2}
    \and Daiping Zhang
      \inst{4}
    \and Liyong Liu\thanks{*Corresponding author}
      \inst{5}
   \and Ali Esamdin\thanks{*Corresponding author}
      \inst{1,2}
}

   \institute{Xinjiang Astronomical Observatory, Chinese Academy of Sciences, 
				Urumqi, 830011, China; {\it guwenbo@xao.ac.cn, aliyi@xao.ac.cn}\\
 		\and
            University of Chinese Academy of Sciences, Beijing, 100049, China\\
             \and
            School of Computer Science, Inner Mongolia University, Hohhot, 010021, China\\
             \and
             Xinjiang University, Urumqi, 830063, China\\
             \and
             National Astronomical Observatories, Chinese Academy of Sciences, Beijing 100101, China; liuly@nao.cas.cn\\
\vs\no
   {\small Received~~2024 December 3; accepted~~2025~~November 26}}
\abstract{Cloud coverage is one of the crucial elements of site testing in astronomy. All-sky camera (ASC) images are beneficial for our research on cloud coverage. In this paper, we propose ASCNet, an innovative model specifically designed for classifying nighttime ASC images collected at the Muztagh-ata site from 2022 March to 2024 June. ASCNet integrates ResNet34 with an ASCModule, which employs Depthwise Dilated Convolution and embeds lightweight Squeeze-and-Excitation attention within its branches to extract fine-grained texture information from the luminance channel. The data set is partitioned by category, with 70\% of images assigned to the training set and 30\% to the test set. The model's performance is assessed by comparing its predictions on the test set with manually annotated labels, yielding a consistency rate of 92.7\%. All evaluation metrics of ASCNet are as follows: Accuracy 92.66\%, Precision 83.26\%, Recall 84.25\%, and F1-Score 83.67\%, and both ablation and comparative experiments demonstrate significant superiority over other models. A confusion matrix is utilized to analyze the differences between manual classification and model classification. The statistical results demonstrate the model's excellent classification performance and its robust generalization ability, illustrating that ASCNet has potential for application in future astronomical image classifications.
}

   \authorrunning{Wang et al}            

   \maketitle

\section{Introduction}           
\label{sect:intro}

When searching for suitable astronomical sites to install optical telescopes, factors such as seeing, cloud coverage, night sky background brightness, precipitable water vapor and astronomy meteorological parameters (air temperature, relative humidity, wind speed and direction, etc.) must be considered \citep{2023MNRAS.525.3236Z,2011PASP..123.1334V}. Due to the scattering and absorption of clouds, the celestial light reaching the terminal of the ground-based telescope through optical observing is affected. So, among these elements, cloud coverage is the most straightforward one influencing the telescope's observable time and continuous observation time \citep{2005MeApp..12...77G,2006Sarazin}. At present, statistical and analytical methods for assessing cloud coverage at telescope sites primarily rely on satellite data analysis \citep{2011MNRASCavazzani,2020Lei,2022MNRAS.511.5363W,2020RAA....20...81C} and ground-based all-sky camera monitoring \citep{2008SPIE.7012E..26M}. A common ground-based cloud detection device is all-sky camera (hereafter ASC; \citeauthor{amt-15-3629-2022} \citeyear{amt-15-3629-2022}), which can provide continuous and unambiguous cloud coverage and distribution information, unaffected by fixed location constraints. ASC is mainly composed of Charge Coupled Device and a fisheye lens. This setup enables high temporal and spatial resolution for capturing cloud coverage and its variation characteristics.\par

The Muztagh-ata site is located in the eastern Pamirs Plateau in Xinjiang, China. The geographic coordinates of the main monitoring area are approximately $38^\circ 21'$ north latitude and $74^\circ 54'$ east longitude, with an altitude of about 4500 meters \citep{2022PASP..134a5006X}. It is a typical high-altitude climate with frigid temperatures, drought conditions and sparse vegetation. A lot of research on optical observation conditions has been done at the site, such as ground meteorology and sky brightness \citep{2020RAA....20...86X}, seeing conditions \citep{2020RAA....20...87X}, surface temperature inversion \citep{2020RAA....20...88X,2025Atmos..16..897Z}, precipitable water vapor \citep{2022PASP..134a5006X}, night-time cloud statistics and continuous observing time \citep{TTWL202304015,2025MNRAS.541.3353G}, the fragmentation of observing time \citep{2024RAA....24c5003G}.\par

With the development of deep learning models in computer vision, the proposals of models such as ResNet\citep{7780459} and Transformer\citep{2020arXiv201011929D,9710580} has promoted advancements in the field of image classification. Based on this background, researchers have attempted to apply these deep learning methods to cloud image analysis and have achieved promising results \citep{JGHW202312001}. \citet{2008Cazorla} analyzed cloud cover estimation and characterization using a method based on an optimized neural network classification procedure. According to \citet{QXKJ201706009}, a method based on image enhancement and image restoration was proposed to research ground-based visible ASC images, significantly enhancing the image quality, especially for thin clouds and low visibility clouds. \citet{1022756519.nh} developed an automatic classification of ASC images based on CATE and traditional machine learning algorithms. According to \citet{TWJZ202402011}, a method adopted CWFF (channel weighting-feature fusion) structure to calculate cloud cover during the daytime, improving the overall accuracy rate and reducing the average absolute error. \citet{YTWT201902012} improved the AlexNet model for ASC images classification, achieving excellent application results. According to \citet{2020RemS...12..464L}, a novel method named Multi-Evidence and Multi-Modal Fusion Network, designed to achieve extended cloud information by integrating heterogeneous features within a unified framework.\par

Despite certain progress in cloud image analysis, existing methods often struggle to balance global semantic understanding and local luminance texture extraction, especially for nighttime ASC images. To address this challenge, we propose ASCNet, a novel classification framework specifically designed for nighttime ASC images, which integrates complementary feature extraction, attention mechanisms, and an optimized training strategy. The main innovations of this paper are outlined as follows:\par
(1) An improved ASCModule is proposed, which integrates depthwise dilated convolution with a lightweight Squeeze-and-Excitation (SE) attention block embedded in each branch to enhance the extraction of fine-grained luminance texture features.\par
(2) A complementary RGB–luminance feature extraction structure is proposed, in which ResNet captures global semantic information while the ASCModule focuses on local luminance textures. This design enables synergistic enhancement between the two feature types and balances global discrimination with local detail representation.\par
(3) A freezing strategy for the ResNetBackbone is designed to improve training stability and generalization capability, making the model particularly suitable for the Muztagh-ata dataset.\par
(4) A lightweight Efficient Channel Attention (ECA) module is introduced in the feature fusion stage to model inter-channel dependencies without dimensionality reduction, thereby improving the efficiency of multimodal feature interaction.\par

The rest of the paper is structured as follows. In Section 2, the data acquisition and data processing methods are given. In Section 3, the basic machine learning model, ASCNet, is described. In Section 4, experiment procedure and implementation details are shown. In Section 5, output display of model labels compared with manual labels and model evaluation results are given. In Section 6, the conclusion is presented.

\section{Datasets and Classification Criteria}

The images in this paper are derived from an ASC employed at the Muztagh-ata site which is composed of a Canon Digital Single Lens Reflex Camera with 3456 × 5184 pixels. The instrument captures ASC images at intervals of approximately 10–20 minutes during daytime and 2–5 minutes during nighttime. In this study, only nighttime images are selected for analysis. The distinction between day and night is made based on astronomical twilight.

\subsection{Image Preprocessing}

\begin{figure}
   \centering
   \begin{minipage}[b]{0.45\textwidth}
       \centering
       \text{(a) Date: July 25, 2023, 00:55:39}
       \includegraphics[width=\textwidth]{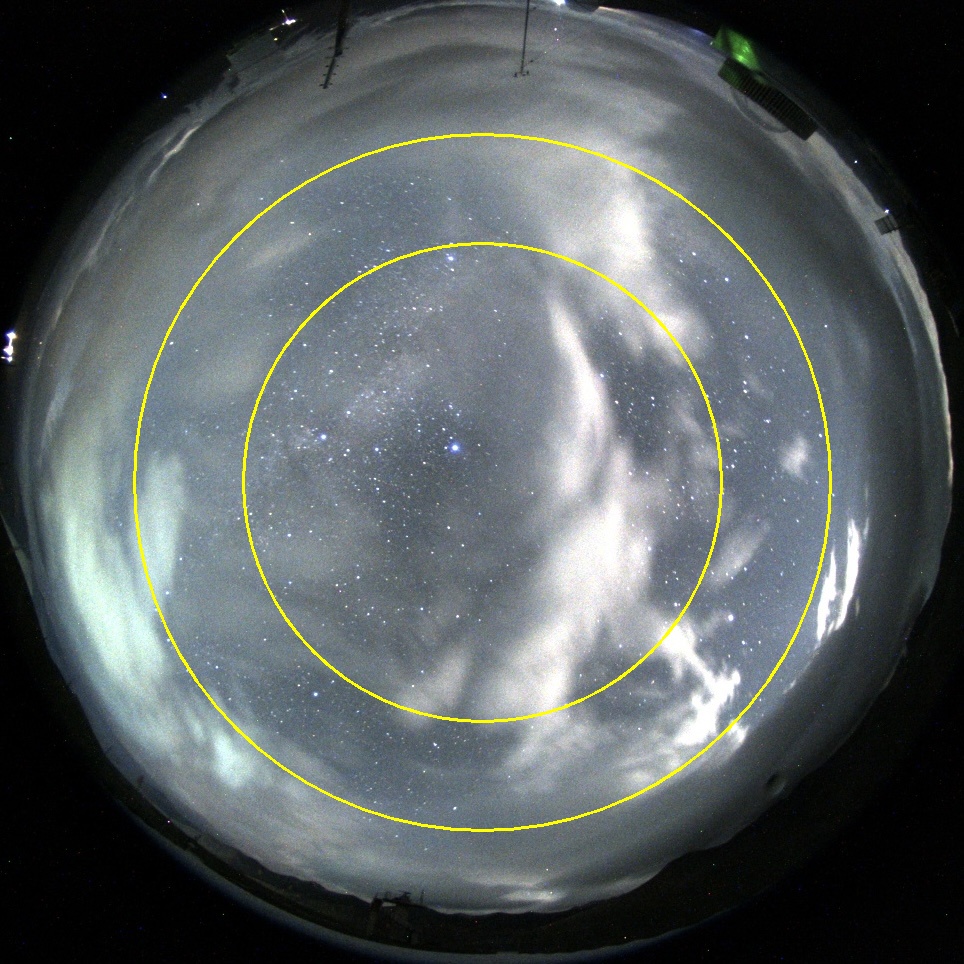}
   \end{minipage}
   \begin{minipage}[b]{0.45\textwidth}
       \centering
       \text{(b) Date: July 25, 2023, 00:55:39}
       \includegraphics[width=\textwidth]{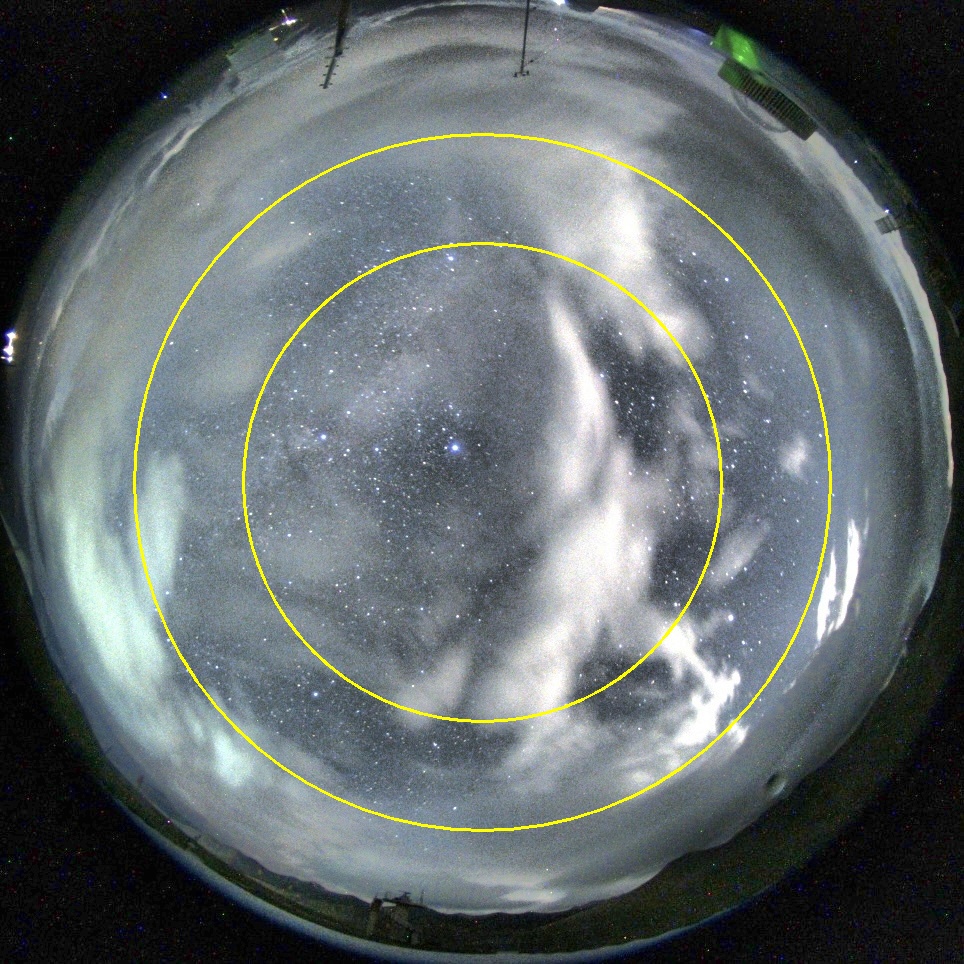}
   \end{minipage}
   \caption{(a) Original image; (b) The pre-processed image.}
   \label{10}
\end{figure}

 Since the ASC is fixed, the images captured by this instrument have consistent imaging dimensions and bit depth, allowing certain preprocessing steps to be omitted prior to model training. However, the images have some extra black edges which require uniform tailoring. Moreover, nighttime ASC images exhibit low overall brightness, as shown in Figure \ref{10} (a). To address the characteristics of such low-light images, Gamma correction is applied to improve the visibility of darker regions, and CLAHE is employed to enhance image contrast. Region segmentation is then performed based on image brightness, and different enhancement strategies are applied to the corresponding segmented regions. This method enhances image feature extraction and highlights the details of thin cloud areas. The enhanced image obtained through this process is shown in Figure \ref{10} (b).\par

\subsection{All-sky Image Classification Criteria}

\begin{figure}
   \centering
   \begin{minipage}[b]{0.3\textwidth}
       \centering
       \text{(a) Date: March 24, 2022, 00:00:42}
       \includegraphics[width=\textwidth]{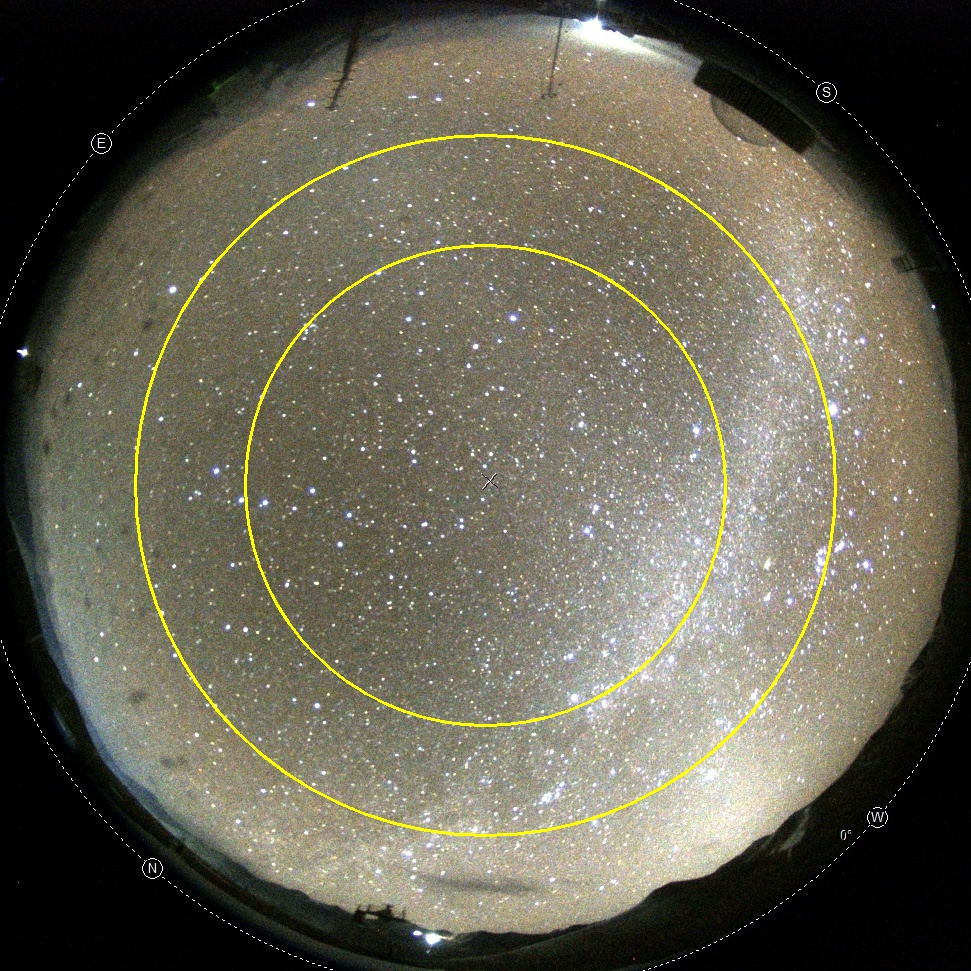}
   \end{minipage}
   \begin{minipage}[b]{0.3\textwidth}
       \centering
       \text{(b) Date: November 29, 2022, 02:58:11}
       \includegraphics[width=\textwidth]{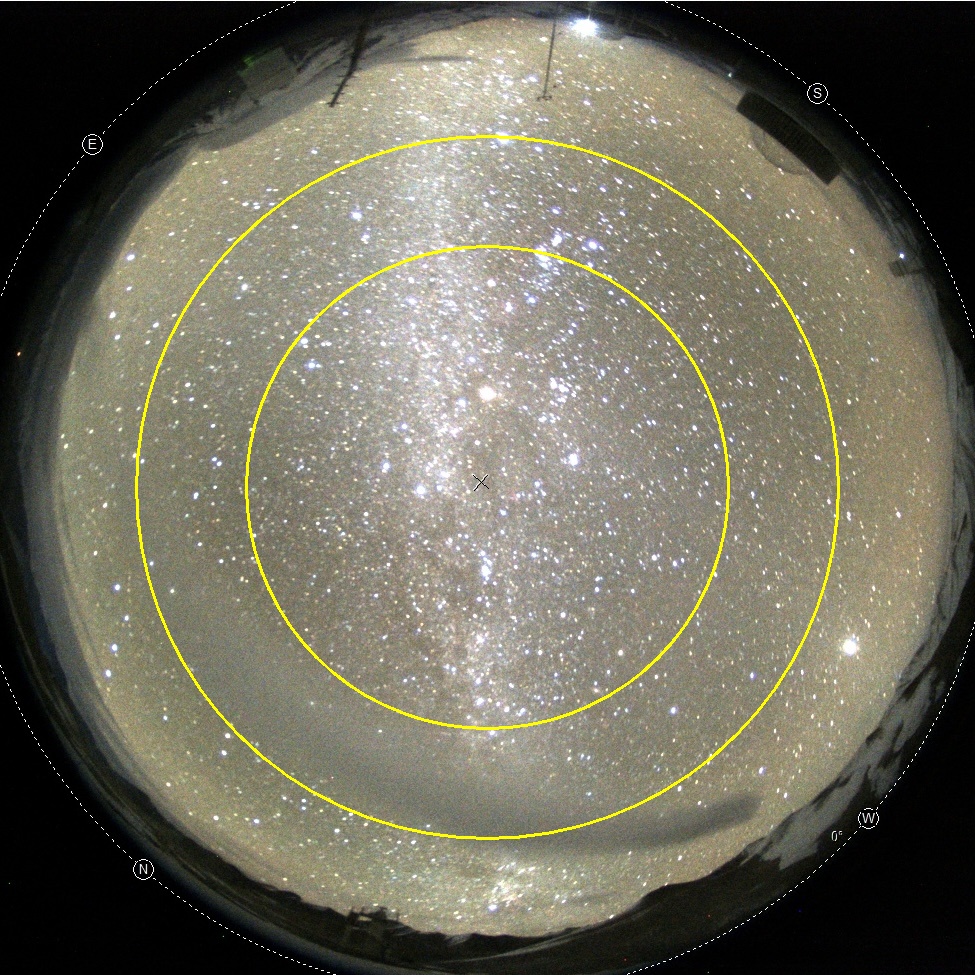}
   \end{minipage}
   \begin{minipage}[b]{0.3\textwidth}
       \centering
       \text{(c) Date: April 17, 2023, 05:02:11}
       \includegraphics[width=\textwidth]{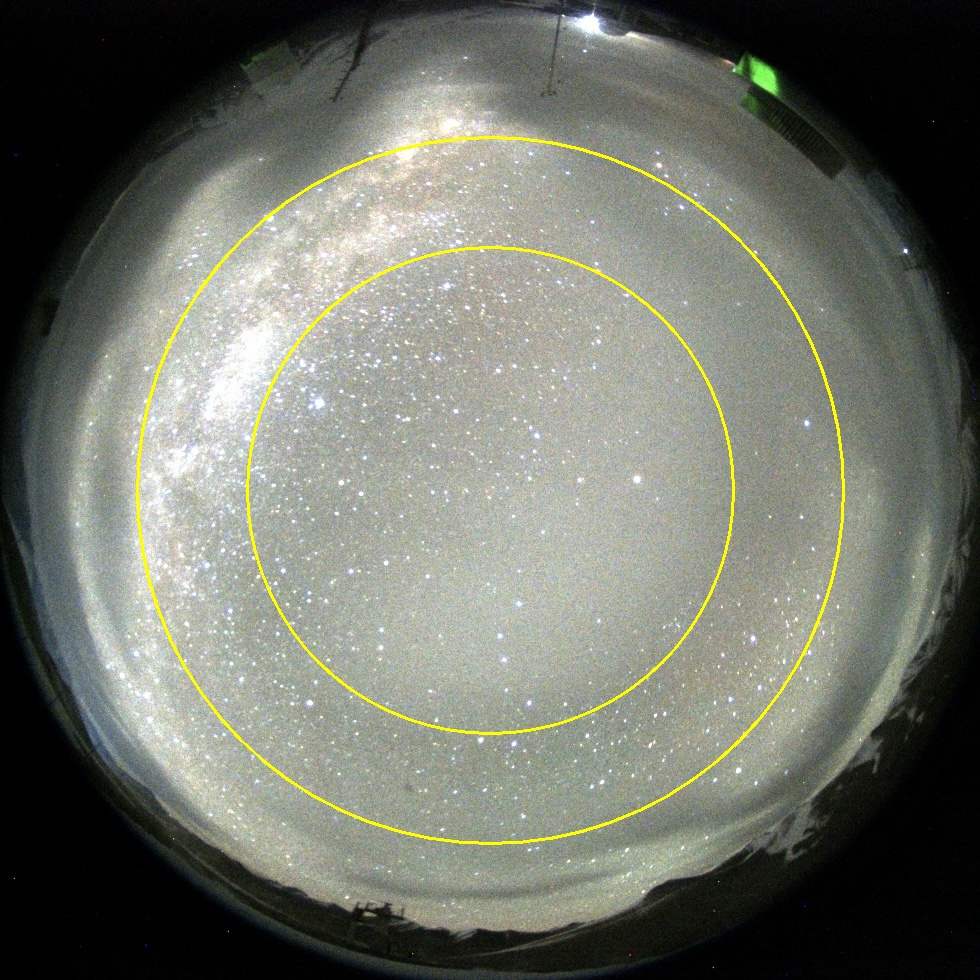}
   \end{minipage}
   \begin{minipage}[b]{0.3\textwidth}
       \centering
       \text{(d) Date: July 24, 2023, 02:17:06}
       \includegraphics[width=\textwidth]{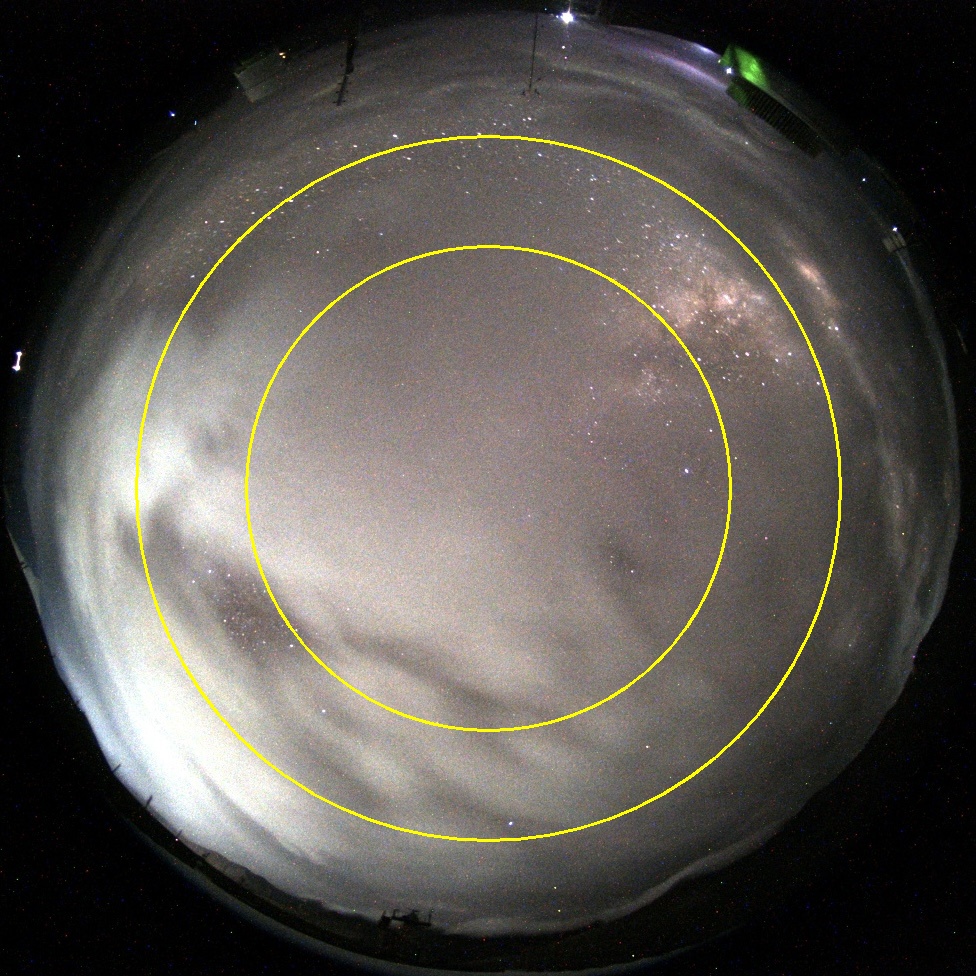}
   \end{minipage}
   \begin{minipage}[b]{0.3\textwidth}
       \centering
       \text{(e) Date: April 14, 2022, 04:52:11}
       \includegraphics[width=\textwidth]{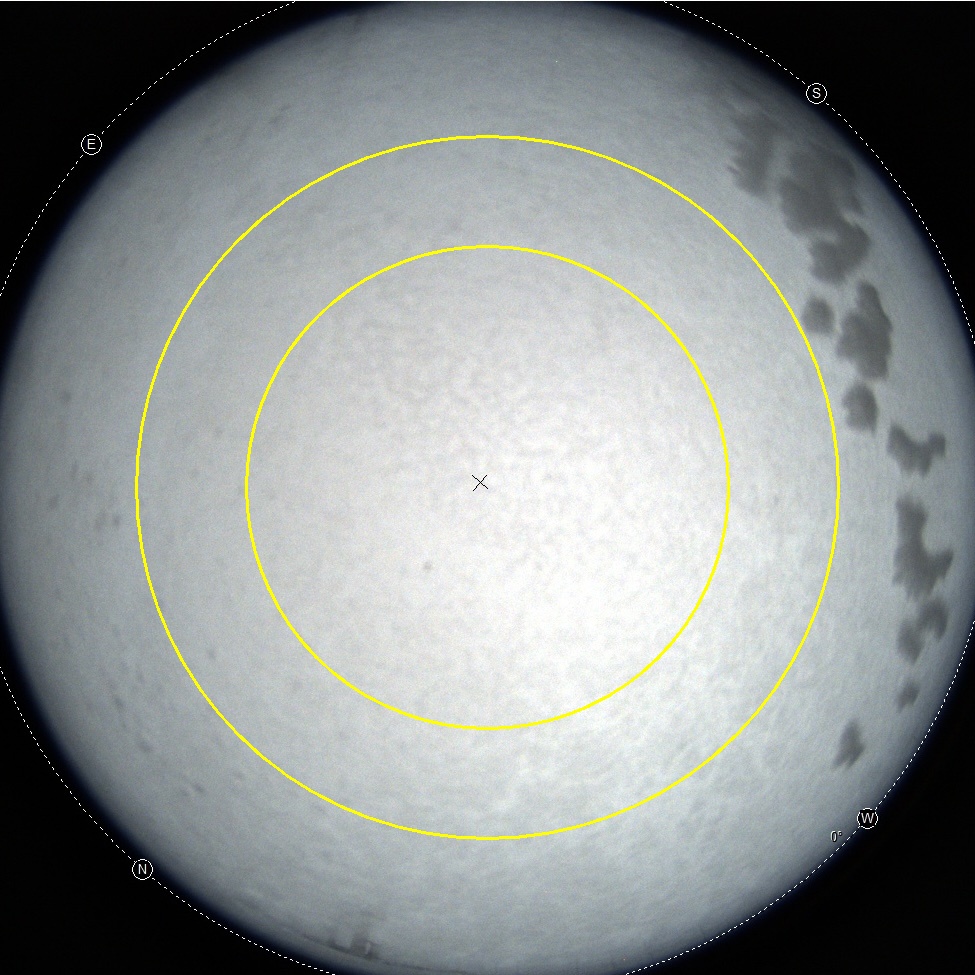}
   \end{minipage}
   \caption{Five categories of ASC images, (a) Clear; (b) Outer; (c) Inner; (d) Covered; (e) None.}
   \label{1}
\end{figure}

The classification of ASC images is conducted according to Skidmore's criteria \citep{2008SPIE.7012E..24S}. The inner circle is stipulated to overlay zenith distance 0°-44.7°, and the outer circle overlays annulus 44.7°-65°. In addition, Skidmore provided seven types of classification, while this paper makes a slight adjustment that give five types as shown in Table \ref{Table 1}, and living examples of these five categories exhibited in Figure \ref{1}. ASC images from 2022 to 2024 were manually classified according to the specified classification standard and a multi-person cross-checking approach was employed to ensure the precision of the results. Through this cross-checking process, manual labels were obtained, representing the classification outcome for each ASC image. From 2022 March to 2024 June, over 150,000 ASC images were acquired at the Muztagh-ata site.  The percentage distribution of manual labels for each category is calculated and presented in Table \ref{Table 2}.\par

\begin{table}
\begin{center}
\caption[]{ Definition of Five Classification Types of ASC Images}\label{Table 1}
 \begin{tabular}{clcl}
  \hline\noalign{\smallskip}
Classification  &  Definition\\
  \hline\noalign{\smallskip}
Clear    &  There is no cloud in the inner and outer circles under optical photography.\\ 
Outer    &  There is no cloud in the inner circle but cloud in the outer circle.\\
Inner    &  There is no more than 50\% cloud in both circles.\\
Covered  &  There is more than 50\% cloud in both circles.\\
None     &  The weather like condensation or snowfall make it difficult to judge cloud cover.\\
  \noalign{\smallskip}\hline
\end{tabular}
\end{center}
\end{table}

\begin{table}
\centering
\caption[]{ Statistics of the Classification of Nighttime All-sky Camera Images at the Muztagh-ata Site from 2022 March to 2024 June}\label{Table 2}
 \begin{tabular}{c c c c c c c}
  \hline\noalign{\smallskip}
Year &  Clear &  Outer &  Inner &  Covered &  None &  Number of images\\
  \hline\noalign{\smallskip}
2022  &  51\%  &  8\% &  7\% &  29\% &  5\% &  21817\\ 
2023  &  54\%  &  8\% &  7\% &  28\% &  3\% &  81279\\
2024  &  54\%  &  2\% &  7\% &  35\% &  2\% &  48057\\
  \noalign{\smallskip}\hline
\end{tabular}
\end{table}

\section{Structure of ASCNet}

\begin{figure}
   \centering
   \includegraphics[width=\textwidth, angle=0]{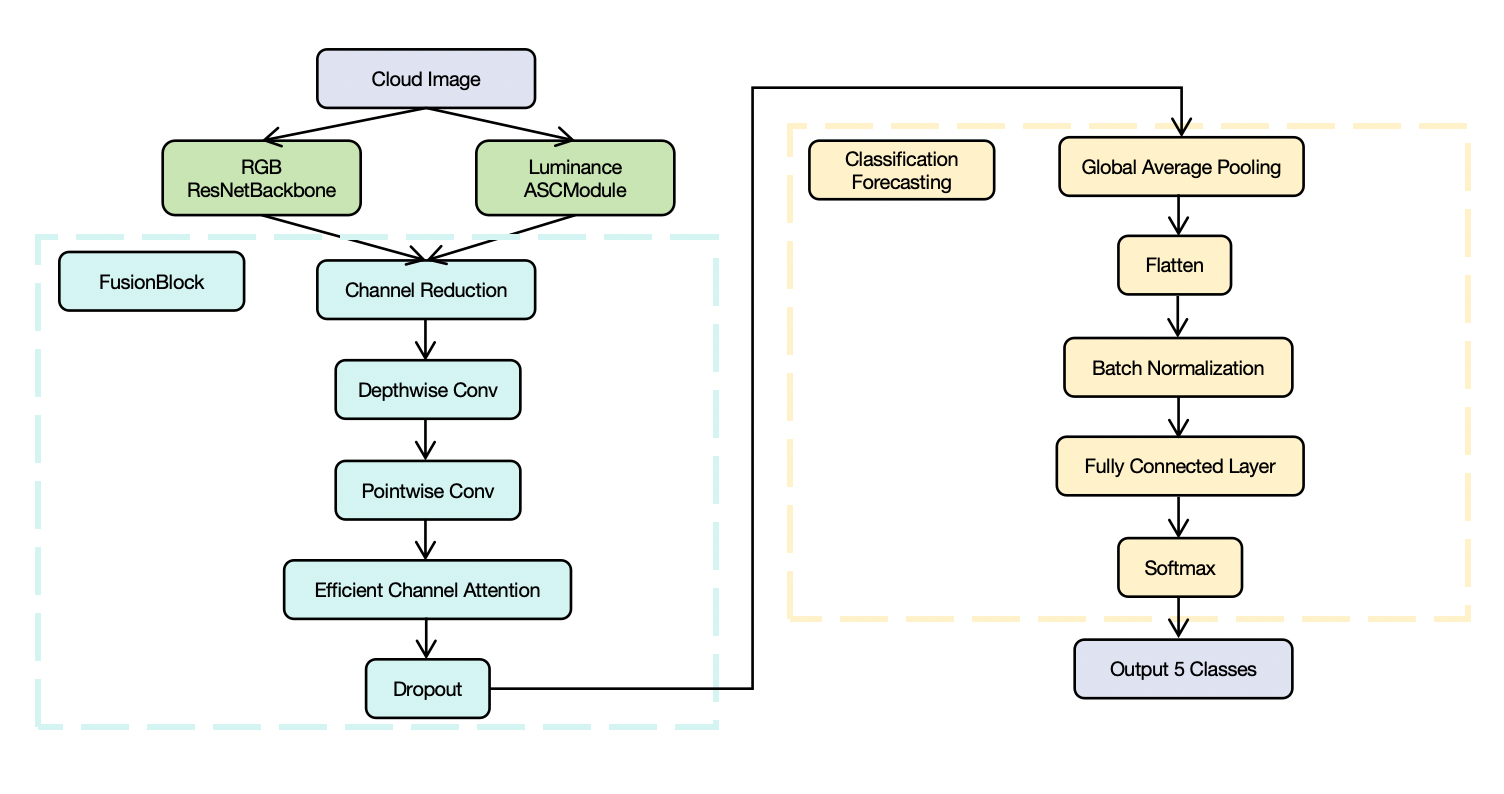}
   \caption{Overall structure of ASCNet.}
   \label{2}
   \end{figure}

The overall architecture of ASCNet is illustrated in Figure\ref{2} \citep{2020E3SWC.18502006W}. Given a color input image $X \in \mathbb{R}^{B\times 3\times H\times W}$, a luminance map $Y \in \mathbb{R}^{B\times 1\times H\times W}$ is first computed using the standard luminance conversion formula \citep{Gonzalez2002}:
$$
Y = 0.299 X_R + 0.587 X_G + 0.114 X_B
$$
The input is then processed through two parallel branches. The RGB branch employs a ResNetBackbone to extract high-level semantic features, while the luminance branch utilizes the ASCModule to capture fine-grained texture features. The outputs of these two branches are subsequently fused via a FusionBlock, which performs channel dimensionality reduction, residual enhancement based on depthwise separable convolutions, and integration of ECA. Finally, the fused feature maps are processed by global average pooling, flattening, batch normalization, and a fully connected layer, followed by a softmax function to generate the final class predictions.\par 

\subsection{ResNetBackbone} 
The RGB branch adopts the ResNet34 as the backbone feature extractor, with the output taken from Layer4, which provides a large receptive field and rich high-level semantic information. The specific parameter information is described in Table \ref{Table 6} \citep{rs14163978}, where the number in parentheses indicates the kernel size. To balance training efficiency and generalization capability, this work supports a freezing strategy: during the initial training phase, several early layers of the backbone are frozen and subsequently unfrozen after convergence, in order to reduce gradient fluctuations and stabilize low-level feature representations. The output feature maps of the backbone are denoted as:
$$
F_{rgb} = \mathcal{B}(X) \in \mathbb{R}^{B\times C_{rgb}\times H'\times W'}
$$
where in this work $C_{rgb}$ is set to 512 for the ResNet34 version.\par

\begin{table}[h]
\centering
\caption[]{Details of ResNetBackbone}\label{Table 6}
\begin{tabular}{c c c c c}
  \hline\noalign{\smallskip}
Layer Name & Output Size & Output Channels & Downsample & Layers\\
  \hline\noalign{\smallskip}
Conv (7×7) & 256×256 & 64 & 2× &1 \\
MaxPool (3×3) & 128×128 & 64 & 2× & 1 \\
Layer1 Block & 128×128 & 64 & 1× & 3 \\
Layer2 Block & 64×64 & 128 & 2× & 4 \\
Layer3 Block & 32×32 & 256 & 2× & 6 \\
Layer4 Block & 16×16 & 512 & 2× & 3 \\
  \noalign{\smallskip}\hline
\end{tabular}
\end{table}

The network body of ResNet34 is composed of multiple residual blocks, each of which is composed of different convolution layers. The following formulas explain how residual blocks work in mathematical notation \citep{DZCL202323014}. Suppose that the output feature of the $l$th residual block is $y_1$, which yields:\par

\begin{equation}
	y_l = h(x_l) + F(x_l, w_l)
\end{equation}
\begin{equation}
	x_{l+1}= f(y_l)
\end{equation}
where $x_l$ is the input of the $l$th residual unit, $F$ is the calculation process of residual unit, $w_l$ = \{ $w_{l,k} $$\mid 1 \leq k \leq K $\} is the series weight of the $l$th residual unit, h($x_l$) = $x_l$ is the shortcut connection, $f$ is the activation function.\par

Ignore the effect of the activation layer, for any $l$th layer there is:\par

\begin{equation}
	x_{l+1} = f(y_l) = h(x_l) + F(x_l, w_l) = x_l + F(x_l, w_l)
\end{equation}
\begin{equation}
	X_{L} = x_{l} + \sum_{i=1}^{L-1} F(x_{l}, w_{i}) = x_{0} + \sum_{i=0}^{L-1} F(x_{i}, w_{i})
\end{equation}

The back propagation chain rule is shown in the equation:\par
\begin{equation}
	\frac{\partial \epsilon}{\partial x_{l}}
 = \frac{\partial \epsilon}{\partial X_{L}} \cdot \frac{\partial X_{L}}{\partial x_{l}}
 = \frac{\partial \epsilon}{\partial X_{L}}(1+\frac{\partial}{\partial x_{l}} \sum_{i=1}^{L-1} F(x_{i}, w_{i}))
 =\frac{\partial \epsilon}{\partial X_{L}}+\frac{\partial \epsilon}{\partial X_{L}} \cdot \frac{\partial}{\partial x_{i}} \sum_{i=1}^{L-1} F(x_{i}, w_{i}))
\end{equation}
where $\epsilon$ denotes loss function that measuring the difference between model predictions and actual labels. Calculating the \(\frac{\partial \epsilon}{\partial x_{l}}\) can guide updated direction and amplitude of parameters. Effectively extracting complex feature information and transferring information from the deep layer L to the preceding layer L-1 is achieved through the use of residual connections. This allows information to be passed between layers,ensuring that the sum term is not constantly equal to -1, thus preventing the problem of gradient disappearance in the residual blocks \citep{WXHK202313037}.\par

\subsection{ASCModule} 
\begin{figure}
   \centering
   \includegraphics[width=0.75\textwidth, angle=0]{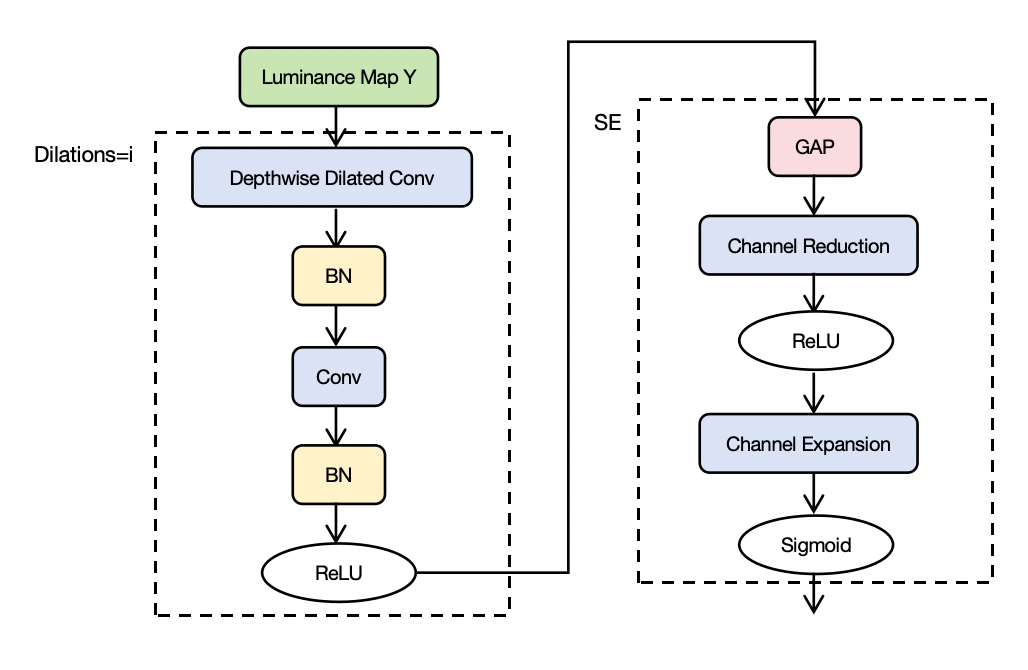}
   \caption{The structure of ASCModule.}
   \label{8}
   \end{figure}

ASC classification is considered a fine-grained task that captures richer low-level features than ResNet34 alone. In this work, the proposed ASCModule employs three parallel branches (with dilation rates i = 1,2,3) as illustrated in Figure \ref{8}, each starting with a convolutional block based on depthwise dilated convolution followed by a lightweight SE attention block embedded within the branch. Compared with traditional parallel convolutions, the use of dilated convolutions \citep{8099558} can expand the receptive field without increasing computational cost, thereby preserving both fine-grained details and contextual information. Specifically, for a luminance map $Y$, the $i$-th branch performs:
$$
F_i = \sigma\big( W_{se}^i \cdot GAP(Conv_{1\times 1}(BN(Conv_{3\times 3, d_i}^{dw}(Y)))) \big) \odot Conv_{1\times 1}(BN(Conv_{3\times 3, d_i}^{dw}(Y)))
$$
where, $d_i$ denotes the dilation of the branch, $GAP(\cdot)$ denotes global average pooling, $W_{se}^i$ denotes the SE channel weights, $\sigma(\cdot)$ denotes the Sigmoid activation function.\par
The outputs of all branches are concatenated along the channel dimension:
$$
F_y = \text{Concat}(F_1, F_2, \dots, F_k) \in \mathbb{R}^{B\times C_y\times H'\times W'}
$$
Moreover, each multi-scale convolution branch of ASCModule is followed by an SE module \citep{8578843}, which adaptively enhances the importance of each channel, and the features from all branches are finally merged and outputted.\par

\subsection{FusionBlock} 
To fully integrate the RGB features and luminance features, in this work we first align their spatial dimensions and then concatenate them along the channel dimension to obtain the combined feature representation $F_{cat}$:
$$
F_{cat} = \text{Concat}(F_{rgb}, F_y)
$$
where $F_{rgb}$ represents RGB semantic features, and $F_y$ represents luminance features. The fused features are then fed into the FusionBlock for further processing.\par
The FusionBlock first applies a 1×1 convolution for channel reduction, compressing the number of channels to an intermediate dimension $C_mid$ to reduce computational cost. It then passes through a depthwise separable convolution residual unit, where the combination of Depthwise and Pointwise convolutions enhances local feature extraction while preserving the original information. Next, a lightweight ECA module captures cross-channel dependencies without dimensionality reduction, improving the interaction efficiency of multimodal features. Finally, Dropout is applied for regularization to enhance the model’s generalization ability. The output of the FusionBlock can be expressed as:
$$
F_{fus} = \text{Dropout}\big(\text{ECA}(\text{ReLU}(BN(DWConv(F_{red}) + F_{red})))\big)
$$
where $F_{red}$ denotes features after dimensionality reduction. This module effectively uses RGB and luminance information, enhancing the model’s ability to represent fine-grained textures and global contextual information.

\section{Experiment Settings}

In line with common practice in large-scale image classification benchmarks \citep{2019Predicting,ju2024monicabenchmarkinglongtailedmedical,2021arXiv211014795Y}, the training set is substantially larger than the test set. Accordingly, our dataset is split such that 70\% of the images from each category are used for training and the remaining 30\% for testing, with no overlap between the two sets. The detailed numbers for each category are presented in Table \ref{Table 3}. This study employs 10-fold cross-validation and averages the results to reduce the impact of random parameter initialization. Compared to a single split validation set, the 10-fold cross-validation can ensure the generalization ability of the model under different data distributions.\par

\begin{table}
\centering
\caption[]{ Number of Images in Training and Test Sets for the Dataset}\label{Table 3}
 \begin{tabular}{c c c c}
  \hline\noalign{\smallskip}
Class &  Training &  Testing &  Total\\
  \hline\noalign{\smallskip}
Clear  &  56683  &  24292 &  80975\\
Outer  &  6489  &  2780 &  9269\\
Inner  &  7077  &  3032 &  10109\\
Covered  &  32492  &  13924 &  46416\\
None  &  3609  &  775 &  4384\\
Total  &  106350  &  44803 &  151153\\
  \noalign{\smallskip}\hline
\end{tabular}
\end{table}

\subsection{Pretraining}

From Table \ref{Table 2}, it can be seen that over the three years, the “Clear” and “Covered” classes account for more than 80\% of the samples, significantly higher than the other categories. These two types of images exhibit pronounced differences in the luminance channel, enabling the model to leverage luminance features for rapid differentiation during the early stages of training. In contrast, the minority classes such as “Outer”, “Inner”, and “None” show relatively weak differences in luminance features and rely more on global RGB semantic information for accurate classification. Therefore, considering the characteristics of the data distribution, it is necessary to introduce a targeted feature learning strategy.\par
The training of ASCNet consists of 20 epochs, with the first 5 epochs employing a staged backbone freezing strategy \citep{2018arXiv180106146H,2025arXiv250712269G}. By fully leveraging the pretrained weights of ResNet34 \citep{7780459} while accommodating feature learning for minority classes, this strategy enables fast convergence for the majority classes and fine-grained representation for the minority classes. Specifically, the staged backbone freezing strategy first freezes certain layers of the ResNet34 backbone (including the RGB feature extraction stem and shallow convolutional blocks), keeping these layers stable based on pretrained weights and excluding them from backpropagation. During this warm-up phase, the model primarily relies on the ASCModule to extract fine-grained texture features from the luminance channel, thereby quickly capturing the discriminative patterns of the “Clear” and “Covered” majority classes. Starting from epoch 6, all previously frozen layers of ResNet34 are gradually unfrozen, allowing the entire network to participate in iterative training. In this stage, the model can integrate global RGB semantic information on top of the existing luminance features, improving recognition accuracy for the “Outer”, “Inner”, and “None” minority classes.\par

\subsection{Evaluation Metrics}
To verify the accuracy of the model's classification results, a confusion matrix is introduced and applied to the test set. Confusion matrices are commonly used in machine learning to evaluate the performance of a classification model and to describe the correctness and error types of the model's predictions. It can help us understand the misclassification of the model and improve the model. Accuracy, Precision, Recall, and F1-Score are utilized as evaluation metrics for the model. The definition are shown in the following equation \citep{WHSP202301013}:\par

\begin{equation}
	\text{Accuracy} = \frac{TP + TN}{TP + TN + FP + FN}
\end{equation}
\begin{equation}
	\text{Precision} = \frac{\text{TP}}{\text{TP} + \text{FP}}
\end{equation}
\begin{equation}
	\text{Recall} = \frac{\text{TP}}{\text{TP} + \text{$FN$}}
\end{equation}
\begin{equation}
	\text{F1-Score} = 2 \cdot \frac{\text{Precision} \cdot \text{Recall}}{\text{Precision} + \text{Recall}}
\end{equation}

where TP, FP, TN, FN represent the amount of data correctly classified of positive class into positive class, the amount of data of wrongly classified negative class into positive class, the amount of data of correctly classified negative class into negative class, the amount of data of wrongly classified positive class into negative class. \par

\subsection{Loss Function}

Since the sample proportions of the “Outer” and “Inner” categories are relatively small, directly using the standard Cross-Entropy Loss may result in insufficient learning for these minority classes. To address the class imbalance problem, Focal Loss \citep{lin2018focallossdenseobject} is employed, which reduces the weight of easily classified samples and enhances the influence of hard examples. The Focal Loss can be formulated as:
\begin{equation}
\mathcal{L}_{FL} = - \alpha_t (1 - p_t)^\gamma \left[ y \log(p) + (1-y) \log(1-p) \right]
\end{equation}
where 
\begin{align*}
p_t &= y \cdot p + (1-y) \cdot (1-p) \\
\alpha_t &= y \cdot \alpha + (1-y) \cdot (1-\alpha)
\end{align*}
Here, \(y\) is the ground-truth label, \(p\) is the predicted probability after sigmoid, \(\alpha\) balances class importance, and \(\gamma\) is the focusing parameter to emphasize hard examples.
This approach improves the classification capability of the minority classes (Outer, Inner) and prevents the model from being overly biased toward the majority classes (Clear, Covered, None). By adjusting the loss weights through Focal Loss, the model focuses more on the “Outer” and “Inner” categories during training, thereby improving classification accuracy for these low-proportion samples. This method effectively mitigates the underfitting issue caused by class imbalance, ensuring a more balanced recognition capability across all categories. \par

\section{Classification Results and Analysis}

\begin{figure}
   \centering
   \includegraphics[width=0.75\textwidth, angle=0]{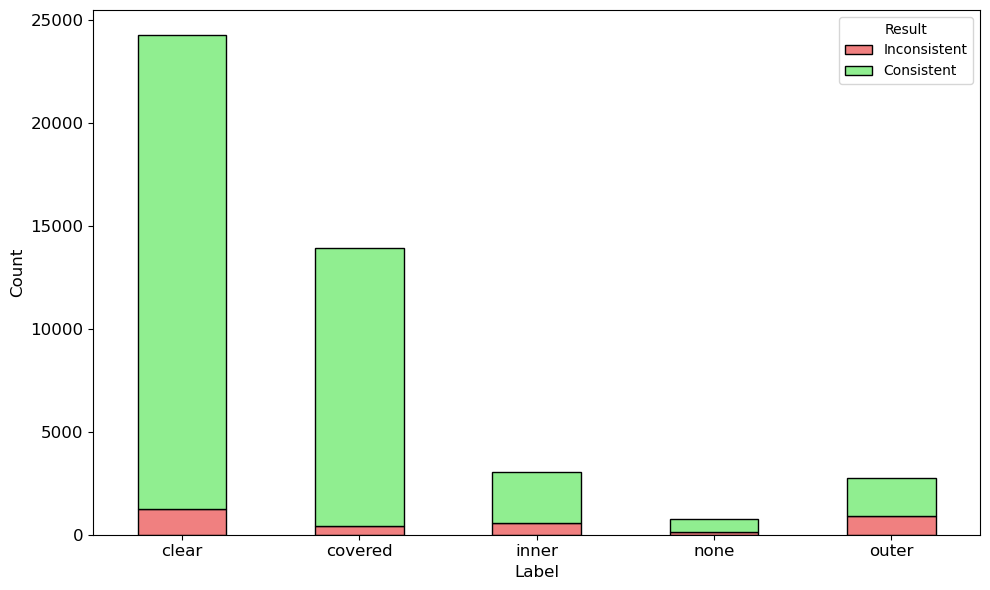}
   \caption{The comparison between the manual labels and the model labels from the test set. In the bar chart, the horizontal coordinate is the five categories of ASC images and the vertical coordinate is the number of occurrences of the categories. Green represents consistency, and red represents inconsistency.}
   \label{3}
   \end{figure}

In order to describe the consistency between the two classification methods, a comparison was made between the manually labeled data and the model data from the test set, as shown in Figure \ref{3}. The proportion of consistency accounted for 92.7\% and the proportion of inconsistency accounted for 7.3\%. It is evident that the model has brilliant generalization performance as shown by the high consistency in the majority of data. The consistency of the classification results from the two methods indicates that ASCNet can be applied to reduce the heavy burden of manual classification in the future. \par

\subsection{Ablation Experiment}

To verify the effectiveness of ASCNet, ablation experiments were conducted to compare the performance of different strategies in ASC image classification \citep{Li2023CloudDenseNet}.
The ablation experiment results are depicted in Table \ref{Table 4} \citep{2018arXiv181201187H}. First, using ResNet34 as the baseline model, the accuracy was only 77.82\%, indicating that relying solely on global RGB semantic features still leaves substantial room for improvement in class discrimination. Subsequently, after introducing the FusionBlock built on ResNet34—specifically, when removing the luminance branch, we adjusted the size of the first module in the FusionBlock to match the output of a single ResNet34—the accuracy increased to 89.94\%, with all evaluation metrics showing significant improvements over the baseline, demonstrating that integrating luminance features can effectively enhance feature discriminability. Further incorporating the ASCModule to construct the complete ASCNet, the Accuracy rose to 92.66\%, and the Precision, Recall, and F1-score all exceeded 83\%, confirming that the ASCModule plays a crucial role in modeling fine-grained texture features from the luminance channel, further compensating for the limitations of using RGB features alone. Overall, each improved module contributes positively to classification performance, validating the rationality and effectiveness of designing the ASCModule based on the characteristics of ASC images.\par

\begin{table}[h]
\centering
\caption[]{Breakdown Effect of Each Component in ASCNet}\label{Table 4}
\begin{tabular}{c c c c c}
  \hline\noalign{\smallskip}
Schemes & Accuracy (\%) & Precision (\%) & Recall (\%) & F1-Score (\%)\\
  \hline\noalign{\smallskip}
ResNet34  & 77.82  & 50.76  & 55.53  & 51.89\\
+Fusion & 89.94  & 77.86  & 78.09  & 74.25\\
+ASCModule+Fusion (ASCNet)    & 92.66  & 83.26  & 84.25  & 83.67\\
  \noalign{\smallskip}\hline
\end{tabular}
\end{table}

\subsection{Comparison Of Different Models}

The experiment compares ASCNet with AlexNet \citep{AlexNet}, DenseNet \citep{DenseNet}, MobileNet \citep{MobileNets}, ViT-B \citep{2020arXiv201011929D}, Swin-T \citep{9710580}, BiFormer \citep{2023arXiv230308810Z}, InceptionNeXt \citep{10657355}. Each model was trained and tested using its officially recommended input resolution: BiFormer and InceptionNeXt used an input size of 224×224, while all other models used 512×512. The classification performance of each model is shown in Table \ref{Table 5}. ASCNet achieves the best overall results, with 92.66\% Accuracy, 83.26\% Precision, 84.25\% Recall, and 83.67\% F1-Score, highlighting its robust and balanced classification performance. In contrast, traditional convolutional networks, such as AlexNet and ResNet34, show notable gaps across all evaluation metrics. Lightweight networks like MobileNet and Transformer-based models (Swin-T, ViT-B) achieve at least a 6\% improvement in Accuracy, but their overall performance still falls short of ASCNet. Recent architectures, including BiFormer and InceptionNeXt, show strong adaptability and better generalization on ASC images. Notably, InceptionNeXt achieves competitive performance with 92.12\% Accuracy and 80.65\% F1-Score, yet ASCNet still surpasses it in Precision and overall consistency. Designed specifically for the characteristics of ASC images, ASCNet demonstrates advantages in dual-channel feature extraction and multi-scale information fusion, resulting in superior performance across evaluation metrics.\par

\begin{table}[h]
\centering
\caption[]{Classification Results of Different Methods on ASC Images}\label{Table 5}
\begin{tabular}{c c c c c}
  \hline\noalign{\smallskip}
Method & Accuracy (\%) & Precision (\%) & Recall (\%) & F1-Score (\%)\\
  \hline\noalign{\smallskip}
ResNet34  & 77.82  & 50.76  & 55.53  & 51.89\\
AlexNet  & 78.71  & 53.20  & 61.71  & 55.65\\
DenseNet  & 84.40  & 53.28  & 57.03  & 54.40\\
MobileNet  & 86.94  & \textbf{79.86}  & 50.43  & 53.45\\
ViT-B    & 84.40  & 53.28  & 57.03  & 54.40\\
Swin-T & 87.63  & 66.67  & 57.89  & 59.52\\
BiFormer & 89.96  & 72.93  & 80.65  & 74.73\\
InceptionNeXt & \textbf{92.12}  & 77.65  & \textbf{86.69}  & \textbf{80.65}\\
ASCNet & \textbf{92.66}  & \textbf{83.26}  & \textbf{84.25}  & \textbf{83.67}\\
  \noalign{\smallskip}\hline
\end{tabular}
\end{table}

\subsection{Confusion Matrix Analysis}

\begin{figure}
   \centering
   \includegraphics[width=0.6\textwidth, angle=0]{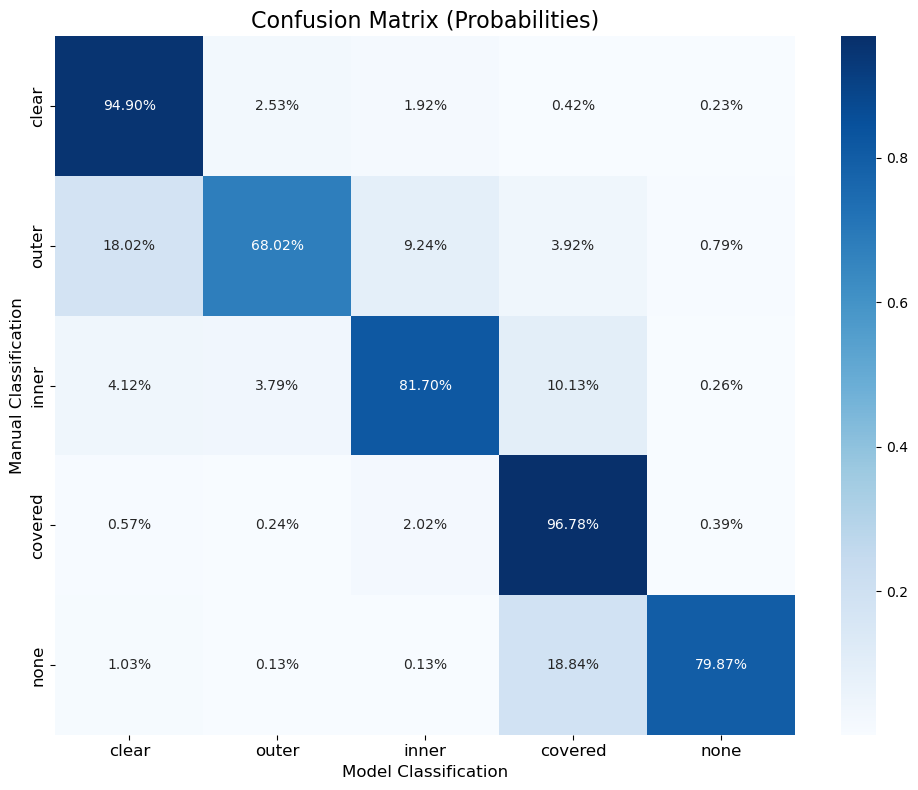}
   \caption{Confusion matrix for text data. The elements on the diagonal of the matrix represent correct classifications, non-diagonal elements represent misjudgements.}
   \label{4}
   \end{figure}

\begin{figure}
   \centering
   \begin{minipage}[b]{0.3\textwidth}
       \centering
       \text{Date: December 08, 2023, 21:33:17}
       \includegraphics[width=\textwidth]{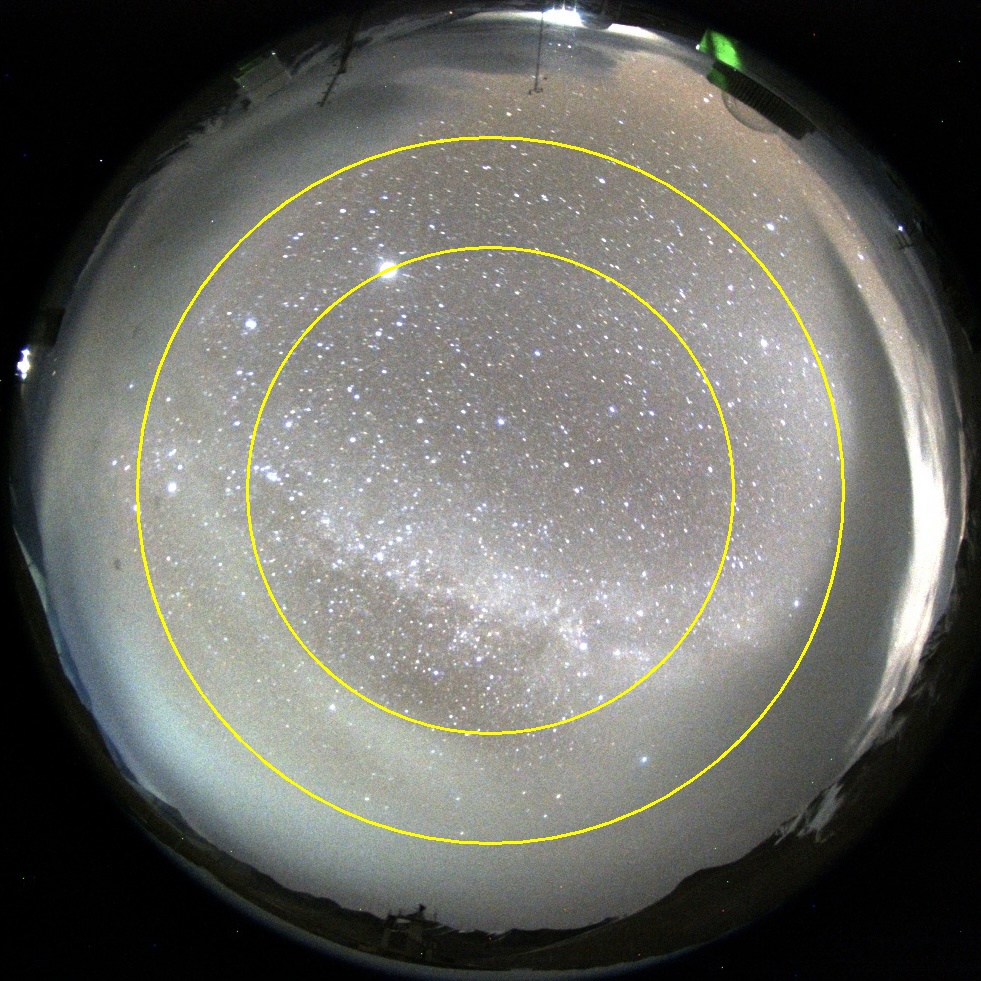}
   \end{minipage}
    \begin{minipage}[b]{0.3\textwidth}
       \centering
       \text{Date:  May 06, 2024, 01:37:14}
       \includegraphics[width=\textwidth]{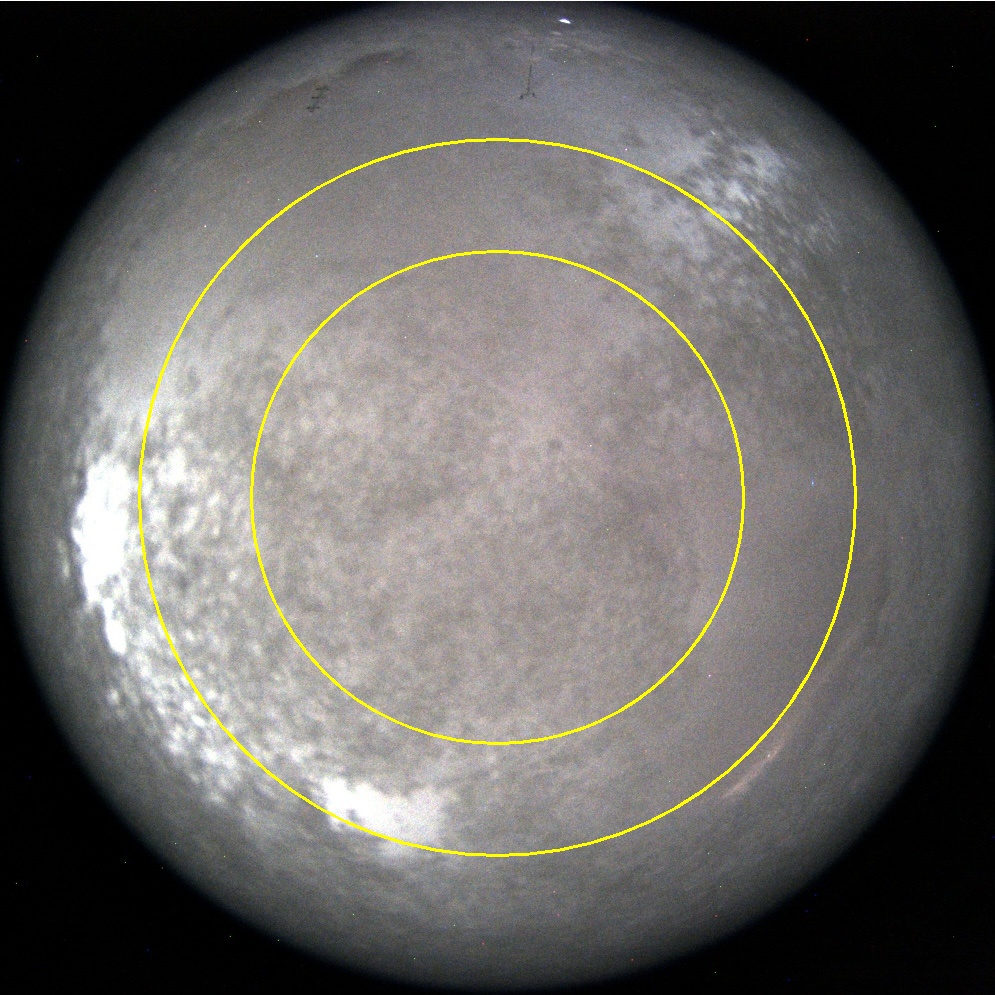}
   \end{minipage}
   \caption{(a) “Outer” is misjudged as “Clear”; (b) “None” is misjudged as “Covered”.}
   \label{5}
\end{figure}

To accurately distinguish the cause of misjudgments, the confusion matrix is further visualized in Figure \ref{4}. The confusion matrix is presented in probabilistic form through normalization. Each row of the matrix represents the true category, and each element in the row represents the proportion of the model's classification results under that category. The elements on the diagonal of the matrix represent correct classifications. The higher the percentage, the better the classification ability. Obviously, the proportion of correct classifications by this model is high. As mentioned earlier, the “Clear” and “Covered” categories account for the majority of the dataset, and both exhibit prominent luminance features. Based on this property, ASCNet is designed to first rapidly learn the feature information of the majority classes, followed by further fine-tuning for the minority classes. However, in practice, it is often challenging for the model to balance all categories simultaneously. It can be observed that the classification performance of the “Outer” class is relatively poor, with an Accuracy of only 68.02\%. Moreover, the “Outer” category is also prone to being misclassified as “Clear” with approximately 18.02\% of images falling into this error. In the annular regions between the inner circle and the outer circle, there may be very few thin clouds. It is difficult for the model to determine whether it is “Clear” or “Outer” in such a similar case. Particularly, a strange point appears in the confusion matrix: instances labeled as “None” are often misclassified as “Covered”. This misclassification may be attributed to the model placing greater emphasis on features characteristic of the Covered category, which can sometimes be present in “None” as well. Typical images of this common misclassifications are shown in Figure \ref{5} \citep{YTWT201902012}.\par

\section{Conclusion}

In this paper, images from the all-sky camera at the Muztagh-ata site were utilized to design ASCNet, an innovative classification model tailored for nighttime ASC images. ASCNet integrates ResNet34 with an ASCModule to achieve complementary RGB–luminance feature extraction, combining global semantic information with fine-grained luminance textures for accurate nighttime ASC image classification. Following common practice in large-scale image classification, the dataset was split for each category with a 7:3 ratio. After training ASCNet with the training set, its feasibility was verified on the test set based on manual labels. The 92.7\% consistency displayed in the bar chart demonstrates that ASCNet, designed based on ASC images collected at the Muztagh-ata site, achieves outstanding classification performance. Four evaluation metrics were introduced to assess the model’s performance, with Accuracy and Precision reaching 92.66\% and 83.26\%, respectively. The confusion matrix visually analyzes two common misclassification situations: “Outer” is misjudged as “Clear” and “None” is misjudged as “Covered”. Obviously, ASCNet has strong generalization performance, and both ablation and comparative experiments verify that it achieves higher classification accuracy compared to other models. This machine learning method can come into use to reduce the workload of manual classification and output objective results. Furthermore, we use the general structure of ASCNet which can be optimized to chase faster classifying speed. The model is applied at the Muztagh-ata site and can be put into use at more observatory sites in the future.

\begin{acknowledgements}
This work is supported by the Chinese Academy of Sciences (CAS) “Light of West China” Program (grant No. 2022 XBONXZ014), the National Natural Science Foundation of China (grant No. U2031209), and Tianshan Talent Training Program (grant No. 2023TSYCLJ0053).
\end{acknowledgements}

\bibliographystyle{raa}
\bibliography{reference}  

\end{document}